\documentstyle[epsfig]{aipproc}

\begin{document}
\title{Dynamics on the Edge: Star-Disk Interaction}

\author{James M. Stone$^*$}
\address{$^*$Department of Astronomy, University of Maryland, College Park, MD 20742}

\maketitle

\begin{abstract}
Our understanding of the dynamical processes which control the
structure and evolution of the interaction region between an
accretion disk and the central star is reviewed.  If the central
star is unmagnetized, this interaction is in the form of a
classical boundary layer.  However, if the central star
is strongly magnetized, it is expected that the inner disk is truncated
by the stellar field, and the accretion flow follows the stellar
field lines to the magnetic poles.  Several outstanding questions
remain regarding this picture.  It is likely that numerical MHD
simulations will prove essential to answering these questions.
However, in order to model the MHD turbulence, angular momentum
transport, and possible dynamo action associated with the
dynamics of star-disk interaction, fully three-dimensional
simulations spanning long dynamical times are required.
\end{abstract}

\section{Introduction}

There are a great number of outstanding and important problems to be
addressed before a detailed understanding of the dynamics of accretion
disks is reached.  One such problem, which recently has been the focus
of increasing interest, is the nature of the interaction between the
inner regions of the disk and the central object.  In the classical
picture of accretion disk structure (e.g. \cite{pringle81}), the inner edge
of the disk terminates in a thin boundary layer at the stellar surface
(unless, of course, the central object is black hole).  In this
boundary layer, the angular velocity of the accreting material
decelerates from near the Keplerian value $\Omega_{K} =
(GM_{*}/R_{*}^{3})^{1/2}$ (where $M_{*}$ and $R_{*}$ are the mass and
radius of the central star), to the stellar rotation rate
$\Omega_{*}$.  If the central star is slowly rotating ($\Omega_{*} \ll
\Omega_{K}$), up to 1/2 of the total accretion luminosity is released
in this boundary layer.  This region is therefore
an important ingredient of the overall
accretion flow: it is where the `rubber meets the road'.

The structure and dynamics of boundary layers have been topics
of investigation for over 20 years.  Three uncertain but important
components determine a model for the boundary layer:
(1) the inner boundary condition for the orbital frequency ($\Omega_{*}$).
(2) the angular momentum transport mechanism in the accretion disk, and
(3) the nature of radiation transport within the disk.  For example,
by adopting the usual ``$\alpha$-prescription" \cite{ss73} for the shear stress
in the disk, it is straightforward to integrate the vertically
averaged disk structure equations to compute a solution for the
boundary layer.  Recently, a succinct study \cite{np93} of how
varying the mass accretion rate and optical depth affects the 
boundary layer has been presented.  The basic conclusion is that
optically thin boundary layers are geometrically thick, a result that is
in many ways related to the existence of the advection dominated
accretion flows within the $\alpha$-viscosity ansatz (because
optically thin boundary layers cannot cool, $\alpha$ is large and
a thick, viscous boundary layer results).

More recently, two-dimensional, time-dependent radiation hydrodynamic
simulations of the optically thin and thick boundary layers have
been presented in \cite{kl96}.  As in \cite{np93}, the angular
momentum transport is specified by use of an $\alpha$-parameter.
Because these models follow the full multi-dimensional dynamics,
the development of very (vertically) thick boundary layers that
engulf the entire star was noted at high mass accretion rates.

\section{MHD Models of the Interaction Region}

In the past five years, it has become evident that magnetic fields
are critical to angular momentum transport within the disk (e.g.,
see Balbus, this volume).  Moreover,
if the central object is strongly magnetized, the structure of
the star-disk interaction region can be completely
changed in comparison to an unmagnetized central star.  No longer
will a boundary layer be formed where the disk surface rubs
against the star.  Instead, the
inner edge of the disk can be disrupted by the magnetic field, and the
resulting accretion flow becomes more complex.  Observations collected
over the past two decades indicate that this is a prevalent phenomena.
For example, it is well established that accretion onto a magnetized
neutron star occurs in X-ray pulsars in binary systems (see, e.g., the
review of \cite{nagase89}).  In some sources, such as Her X-1, SMC
X-1, and Cen X-3, it is clear that the infalling matter is processed
through an accretion disk around the pulsar (\cite{nagase89}; \cite{tjemkes86}).
In the case of accreting
white dwarfs, there is direct observational evidence that the accretion
disk is truncated in magnetized stars: in weakly magnetized systems such as
DQ Her the disk extends inwards to a boundary layer, while in AM Her
systems (in which the white dwarf is known to be strongly magnetized
due to the presence of polarized cyclotron emission), the inner disk
and boundary layer is absent (e.g. see \cite{kylafis80}).  More
recently, sophisticated spectroscopic observations of young stellar
objects (YSOs) which are still accreting indicate that in these systems
as well, the accretion disk is truncated at several stellar radii (most
likely by a magnetic field, \cite{bertout88}), and that the accretion
flow is consistent with free-fall along dipole field lines toward the
polar caps rather than an equatorial boundary layer \cite{hartmann94}.
Moreover, direct observational evidence that these
YSOs are the source of strong magnetic fields comes from the presence
of starspots \cite{bertout88}, and from X-ray and polarized radio
emission (e.g., \cite{montmerle93}).

Theoretically, the interaction of a stellar magnetic field with an
accretion disk has important consequences for the evolution and
observed spectrum of the source.  For example, in the case of accreting
magnetic neutron stars, dynamical processes acting between the disk and
star seem essential to account for the spin-up and spin-down of X-ray
pulsars (\cite{pringle72}; \cite{ghoshlamb}; \cite{lovelace95}).
In the case of accreting white dwarfs, the disruption of the
inner disk and lack of a boundary layer can strongly affect the high
energy spectrum.  Similarly, in the case of YSOs,
the interaction may account for the slow rotation rates of accreting
YSOs \cite{konigl91}, and many aspects of the UV spectrum and
variability (\cite{bertout88}; \cite{konigl91}).  Moreover, the
interaction is often invoked to explain the source of mass outflows
which are observed to be ubiquitous in star forming regions (\cite{edwards93}).
For example, Frank Shu and his
collaborators are developing an intriguing theory for
magnetocentrifugally driven winds which arise from the interaction
region (\cite{shuetal94}, see
also \cite{lovelace95}).  It is important to note that in both
compact objects and YSOs, the magnetosphere-disk interaction produces
more than simply esoteric changes to the accretion flow near the
stellar surface.  Rather, the magnetohydrodynamics (MHD) of the
interaction is thought to be fundamental to the interpretation of
observed stellar rotation rates, the spectrum (which is strongly
affected by the presence
or absence of a boundary layer), and to the production of winds and
outflows.

The theory of the interaction of a stellar magnetosphere and an
accretion disk is still being developed.  In early work, Scharlemann
\cite{sch78} and Aly \cite{aly80} argued that near compact objects, the
conductivity of the accreting plasma is sufficiently high that surface
currents in the disk will screen it from the stellar magnetic field.
Thus, they argue, the stellar magnetic field is pinched inwards until
the disk is truncated at a few stellar radii, and they calculate
analytically the steady-state geometry of the field in this case.
Plasma enters the magnetosphere only via turbulent diffusion
driven by the Kelvin-Helmholtz (K-H) and Rayleigh-Taylor (R-T)
instabilities, which these authors suppose are confined to a thin layer
along the edge of the disk.  On the other hand, in an influential
series of papers, Ghosh \& Lamb (\cite{ghoshlamb} hereafter
GL) argued that these same instabilities should be so effective as to
cause the stellar magnetic field to diffuse rapidly into the disk.  To
aid in this process, they point out that if angular momentum transport
mechanism in the disk is associated with turbulence (as is supposed in
an ``$\alpha$-disk" model), then this turbulence in combination with
reconnection between the disk and magnetospheric fields can increase
the diffusion rate of the field into the disk.  Therefore, GL argue, the
net effect is that the interaction region of the disk which is threaded
by the stellar magnetic field is greatly increased by these processes
in comparison to the ideas of Aly and Scharlemann.

More recently, Shu et al have constructed a detailed theory of the
interaction region in YSO disks.  In their model, the disk is truncated
at a radius $R_{t}$ which is only slightly less than the corotation
point $R_{x} = (GM_{*}/\Omega_{*}^{2})^{1/3}$ of the rigidly rotating
magnetosphere and disk.  A fundamental difference between YSO disks and
those around compact objects is that because the former are so dense and
cold, the fluid is only partially ionized, and microscopic
processes can lead to macroscopic diffusion of the magnetic field.
Still, in the Shu et al model, the magnetospheric field threads the
disk only in the narrow region $R_{t}<r<R_{x}$, in sharp contrast to the
ideas of GL.  An essential ingredient of Shu's model is that beyond
$R_{x}$, the magnetic field lines can drive a magnetocentrifugal wind
(\cite{najita94}, see also \cite{lovelace95}) similar to the disk
winds studied by \cite{wardle}, while below $R_{x}$, material
accretes onto the central star via a polar funnel flow  \cite{ostriker}.
The disk wind opens magnetic field lines beyond $R_{x}$, while
accretion along the field lines below $R_{x}$ implies they are closed,
leading to an ``X-point" geometry for the field at $R_{x}$.  Angular
momentum transport across the ``X-point", which is essential for the
model, is thought to occur via a combination of the Balbus-Hawley (B-H)
instability (\cite{bh1}; \cite{bh2}) and the disk wind mechanism.

\begin{figure}[tb!]
\centerline{\epsfig{file=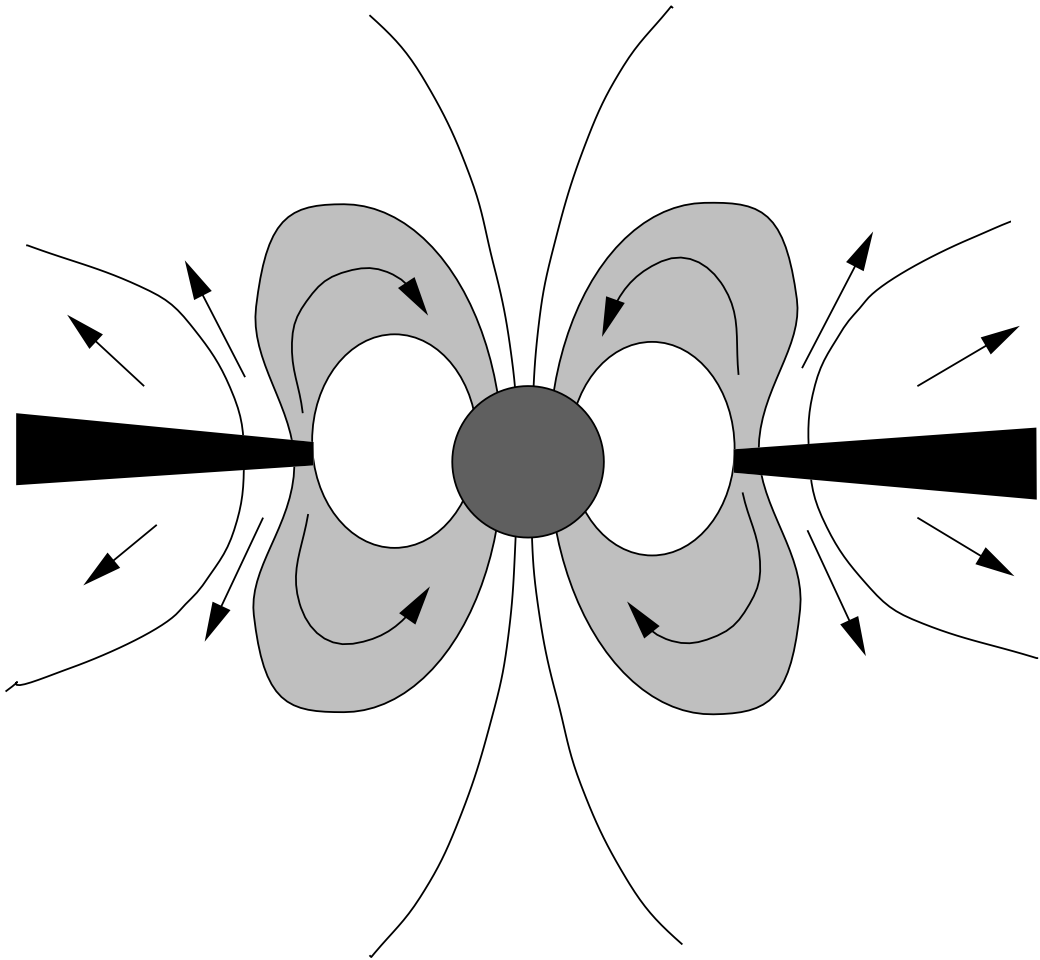,width=5.5in}}
\vspace{10pt}
\caption{A sketch of the basic geometry of an accretion flow near a
strongly magnetized central star.  The disk (black region) is truncated
in the inner regions by the stellar field.  The accretion flow (shaded)
follows the stellar field lines towards the magnetic poles.  Outflow
may occur along open field lines at larger radii.} \label{fig1}
\end{figure}

If one combines the fundamental ingredients of all of these models,
the result can be summarized by the sort of system sketched
in Figure \ref{fig1}.
Rather than being the complete picture, the model summarized by
Figure \ref{fig1} instead raises important questions:
\begin{itemize}
\item {\em What is the width of the region where the stellar magnetic
field lines thread the disk?}
\item {\em What is the role of winds and outflows in controlling
the dynamics of the star-disk interaction region?}
\item {\em What is the three-dimensional structure of the sheared
magnetic field?}
\end{itemize}
Critiques of this picture can also be found in, e.g., \cite{safier98}.

It is quite likely that complete answers to these three questions
will be possible only through the use of numerical methods
to study the MHD  of the interaction region in detail.  For example,
it is important to note the pivotal role that MHD instabilities play in
each theory in determining the basic properties of the interaction
region.  In both YSO disks, and disks around compact objects, the
instabilities control the diffusion rate of the magnetic field into the
disk (and therefore the width of the interaction region), and regulate
fundamental physical processes such as angular momentum transport in
the disk, and the mass accretion rate.  In the theory of Aly, nonaxisymmetric modes of the K-H and R-T
instability load plasma onto the magnetospheric field lines, but only
in a narrow region near the inner edge of the disk.  In the theory of
GL, the K-H instability, as well as turbulent diffusion and
reconnection, result in a much larger interaction region.  In the
theory of Shu et al, the B-H instability plays a role in transporting
angular momentum across the X-point.  In each case, a predictive theory
requires detailed knowledge of the nature of these instabilities in the
nonlinear regime, which in turn will require numerical
simulations.
Simulations also have the added advantage of being able to
compute the global structure of the sheared magnetic field
self-consistently with the dynamics, and they will be the
only way to study time-dependent effects and variability, interesting
problems in their own right.

\section{MHD Simulations of the Interaction Region}

It would indeed be fortunate if all that was required to study
the MHD of the star-disk interaction region was an MHD code
and a computer.  However, there are additional requirements which
limit both the applicability and utility of the results of numerical
investigations.  The first, and probably most serious, results from the fact that
a numerical simulation is the solution to an initial value problem:
it is the time-dependent evolution from a particular initial state
subject to certain boundary conditions.  However, it is difficult to specify
an initial condition that properly represents the star-disk
interaction region.  Starting from any general initial condition
(such as uniform fields, or sub-Keplerian rotation) can lead to
strong transients in the solution that may dominate the resulting
evolution.  One must make reasonable assumptions about the initial
and boundary conditions, and then ideally study the evolution using
many experiments in which these initial
and boundary conditions are varied.

Furthermore, the long term dynamical evolution (which is most likely to
be independent of the initial conditions, assuming the system is able
to `forget' its initial state) is the most interesting.  We
are really interested in the evolution in three-dimensions as well, not only
because real systems are three-dimensional, but more fundamentally, the
nonlinear saturation of MHD instabilities that control the interaction
region is different in two- versus three-dimensions.  For example, in
axisymmetry, the Balbus-Hawley instability saturates as the ``channel"
solution \cite{hb1} composed of two radially streaming
columns of fluid; one moving radially inward with a sub-Keplerian
rotational velocity, and the other moving radially outward with a
super-Keplerian rotational velocity, while in three-dimensions, it
saturates as MHD turbulence.  Computing the
long-term dynamical evolution in three-dimensions is beyond the scope
of current computers.

\subsection{Axisymmetric Simulations}

The most productive way to study the evolution from a variety
of initial conditions is to focus first on axisymmetric problems,
despite the restrictions this imposes.  In fact, a variety of
results have been reported recently from several groups.

Hayashi and collaborators \cite{hayashi96} have considered the evolution of
a pure dipole field anchored in the central star and
a Keplerian disk.  Rotation of the disk quickly winds the field
lines up, and the magnetic pressure associated with the
resulting toroidal magnetic field causes the field lines
to expand away from the star in exactly the manner expected
by the analytic work of \cite{lovelace95} and \cite{lb94}.
Resistivity is added to the simulations
to allow reconnection of field lines.
The expanding magnetic field can then pinch-off near
the star, forming an isolated `plasmoid' of closed field lines that
escapes.  The authors associate these reconnection events
with recently observed X-ray flares in T~Tauri stars, and moreover
suggest the plasmoid might be related to the formation of 
protostellar jets.  Since after reconnection the magnetic
field lines resume something like their initial geometry,
it is possible this process continually repeats, producing a very
time-dependent jet.

Goodson  and collaborators \cite{goodson97} have used a nested mesh code to follow
the evolution of a similar problem over very large spatial scales.
Their smallest grid covers the inner few stellar radii, while
their largest grid extend to a few AU.  Not only do they
observe the time-dependent ejection of plasmoids, as
reported by Hayashi et al, but since they study much
larger scales they report the plasmoids are collimated into jets
on the largest scales.  Some of the latest of these results were
reported at this meeting.

Recently, Kristen Miller and I have performed
simulations of disks which are adiabatic,
axisymmetric, has non-zero resistivity, and is initially in Keplerian
rotation \cite{miller97}.  The magnetosphere is assumed to be initially in
magnetostatic equilibrium, corotating with the central star, and
threaded by one of three different initial magnetic field topologies:
1) a pure dipole field which also threads the disk continuously
everywhere, 2) a dipole field excluded from the disk by surface
currents, and 3) a dipole field continuously threading a disk
superposed with a uniform axial magnetic field.  A number of
exploratory simulations are performed by varying the field strength,
the disk density and inner radius, the magnitude of the resistivity,
and the stellar rotation rate.  These simulations are designed as an
initial study of the magnetohydrodynamics of the interaction region.

\begin{figure}[tb!]
\centerline{\epsfig{file=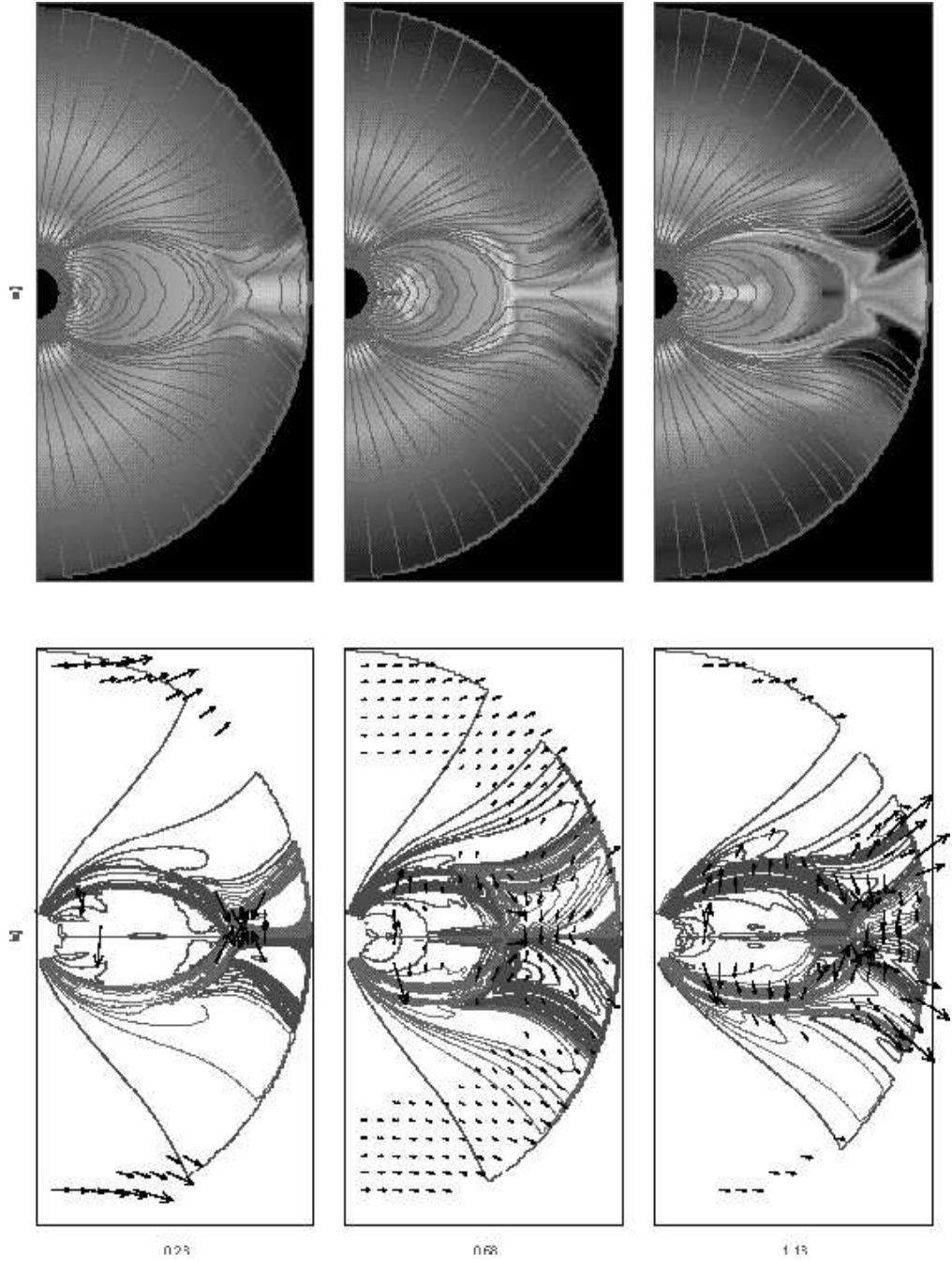,height=7.0in,width=5.25in}}
\vspace{10pt}
\caption{Evolution a star-disk interaction region in which the
field initially consists of dipole plus a constant vertical component.
The top images show poloidal magnetic field lines overlaid on a logarithmic
greyscale image of the density (white represents the highest
densities, black the lowest); the bottom images show
contours of $B_{\phi}$ overlaid on
logarithmic poloidal velocity vectors.}
\label{fig2}
\end{figure}

Generally, we find rapid evolution of the disk occurs due to angular
momentum transport by either the Balbus-Hawley instability or magnetic
braking effects.  Equatorial accretion results on a dynamical
timescale unless the magnetic pressure of the magnetosphere exceeds
the ram pressure of the accreting disk plasma; the latter we find to
be a highly time dependent quantity.  In the case of a pure dipole
magnetospheric field, however, rapid stellar rotation can result in a
field geometry which inhibits polar accretion even when ram and
magnetic pressures balance.  In contrast, we find that polar accretion
can occur regardless of the stellar rotation rate when strong global
disk magnetic fields combine with stellar magnetic fields to create a
favorable net field topology.  Figure \ref{fig2} shows the evolution
in such a topology.  It consists of a
global vertical field combined with a stellar dipole field to
produce a magnetic ``X-point'' topology without the tension forces
characteristic of the pinched dipole geometry (as in, e.g., Figure \ref{fig1}).
This choice was motivated in part by the earlier comprehensive work of
\cite{hirose97}.
We find that this kind
of magnetic ``X-point'' topology produces polar accretion regardless
of the stellar rotation rate (at least for the parameter values
adopted here), suggesting it is ``sturdier'' than the pure dipole
topology.  Thus, the presence of ambient magnetic fields which are
comparable in strength to the stellar field itself (at the inner disk
edge) can promote polar accretion.  This is only true if the disk and
stellar fields are anti-aligned; if they are aligned, the disk field
does not promote polar cap accretion.  If the disk field contains
domains of oppositely directed field, the accretion flow may vary as
each domain is accreted.

Highly time dependent winds are evident in the evolution of all three
field topologies.  The winds are generally channeled along field lines
which have been opened due to reconnection.  The speed and variability
of the outflows is dependent on the magnetic field strength and
accretion topology.  Net torque on the star during accretion is
measured to be positive, i.e., the star is being spun up.

Perhaps the most important aspect of this study
is the fact that the evolution starting from many initial 
conditions was considered, allowing one to assess the ubiquity
of the flow structure sketched in Figure \ref{fig1}.  Not all
simulations resulted in polar cap accretion flows, in fact those
that did were in the minority.  Instead, the controlling aspect
of the evolution was the angular momentum transport mechanism.
Those simulations in which the Balbus-Hawley instability developed
resulted in the magnetosphere being crushed to the surface of the
star because in axisymmetry the nonlinear outcome of the instability
is the exponentially growing channel solution.  However, it is known
from three-dimensional studies that when non-axisymmetric modes
are allowed, the instability drives MHD turbulence.  Modeling this
turbulence in the star-disk interaction region will require fully
three-dimensional simulations.
 
\subsection{Three-Dimensional Simulations}

\begin{figure}[tb!]
\centerline{\epsfig{file=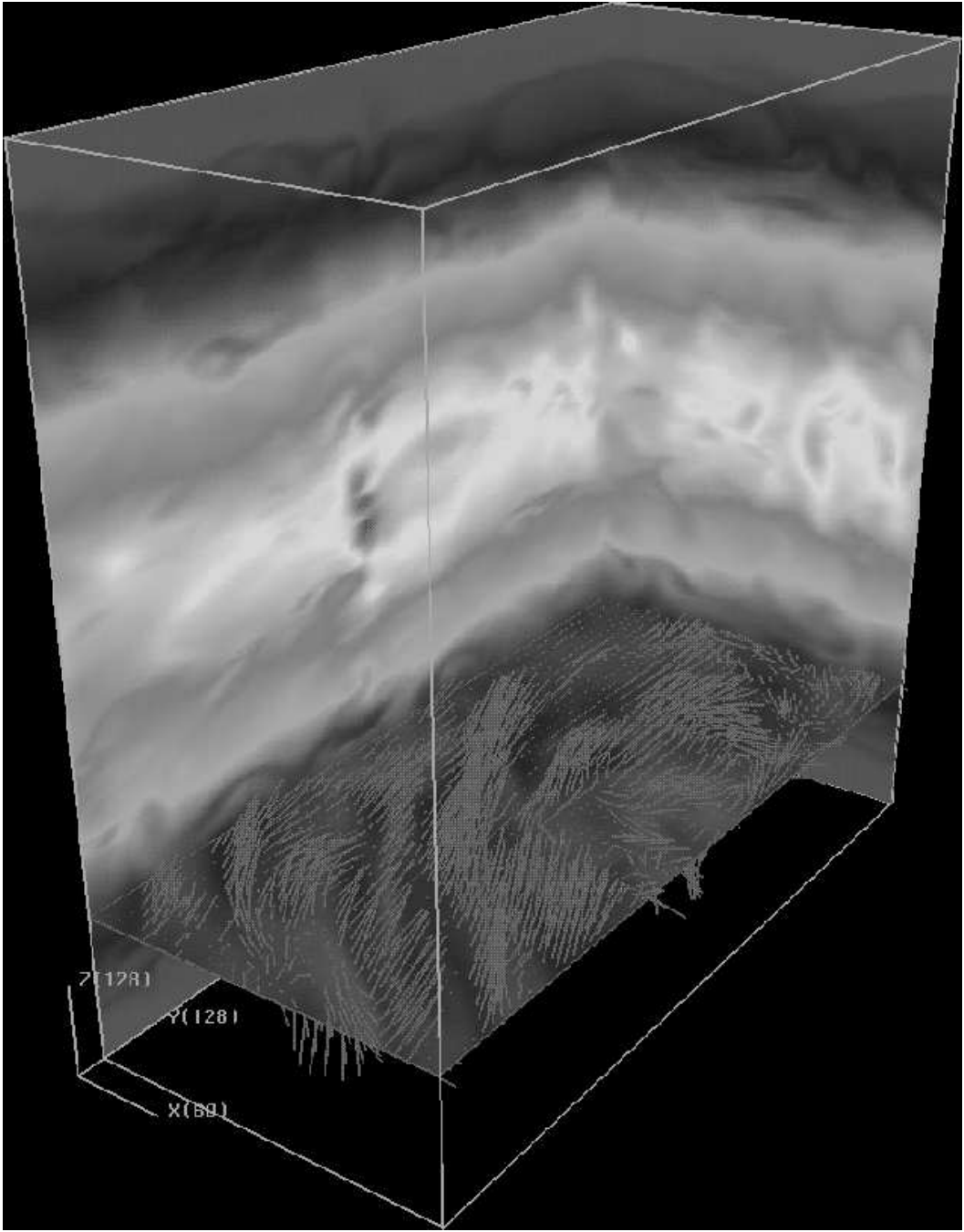,width=5.5in}}
\vspace{10pt}
\caption{Structure produced by the Balbus-Hawley instability in a
local patch of a stratified accretion disk.  The greyscale shows the density
(with white representing the highest densities -- the disk
midplane is orientated horizontally near the center of the box), while
the arrows show the magnetic field through a horizontal slice.  }
\label{fig3}
\end{figure}

Fully three-dimensional simulations of stratified, magnetized accretion
disks designed to study the nonlinear evolution of the Balbus-Hawley
instability have been presented by a number of workers, including
\cite{brandenburg95} and \cite{stoneetal96}.  These simulations
are local, and therefore consider only a small patch of the disk
in the horizontal plane corotating with the disk.
Figure \ref{fig3} shows the
structure which results in a typical model.   There are several important
points to be learned from these simulations.  For example, inspection
of the magnetic field at high-$z$ in the disk (the arrows shown
in Figure \ref{fig3}) clearly demonstrates how complex the field
geometry can become.  Moreover,
the time evolution reveals the field is highly fluctuating as well.
Whether the dynamics of the flow containing such a highly temporally
and spatially varying field can be adequately described by a static
mean field such as shown in Figure \ref{fig1} remains to be examined.

Additionally, the studies of \cite{brandenburg95}, \cite{stoneetal96},
and \cite{hgb2} show that the MHD turbulence generated
by the Balbus-Hawley instability in accretion disks produces
dynamo action in the sense that it strongly amplifies the field
for timescales much longer than the dissipation time.  The power
spectrum of the amplified field follows a declining power law, so
that most of the energy in the field is in the shortest wavenumbers
(longest wavelengths).  The
natural length scale in the disk is the vertical scale height $H$.
For a cold disk, $H \ll R_{*}$, so whether this
field is important to the {\em global} dynamics of the star-disk
interaction region (on scales of order $R_{*}$) is unclear.
If there is an inverse cascade,
perhaps strong field on scales much larger than $H$ can be generated.
However, it seems difficult to understand how fields which span very
large radial domains can exist in a strongly shearing accretion disk.
Moreover, if they do, then most of our accretion disk models
are suspect, since they are based on a the ansatz that the
shear stress is determined by the {\em local} conditions in the
disk.  It is clear that investigations of the global
structure of dynamo generated disk magnetic fields is of great importance.

\section{Summary}

Our understanding of the star-disk interaction region
will undergo rapid progress in the near future as important
developments are being made on both the observational and
theoretical fronts.  For example, observations of 
X-ray binaries and accreting pulsars are providing new
insights into these systems (e.g., see the review of
Bildsten, this volume).  Moreover, infrared and optical
studies of young stars provides constraints from less
extreme environments (e.g., see the review of Calvet, this volume).
Finally, the rapid maturation of computational methods as
a tool to investigate the MHD of the star-disk interaction region
should help theorists to keep up with the observers.

When the central star is not magnetized, the star disk interaction
region consists of a relatively narrow (at least in the radial
direction) boundary layer spread like
a belt where the disk rubs against the star.  On the other hand,
if the central star is strongly magnetized, the star-disk interaction
region is much more complex.  Our current understanding of this
circumstance can be boiled down to a ``working hypothesis" which
is sketched in Figure \ref{fig1}.  To first order, the magnetic
field of the star is taken to be dipolar.  The ram pressure of
accreting material pinches the outer stellar field lines inward,
which generation of toroidal field by shear inflates the stellar
field lines in the axial direction.  Outflows and winds may
be produced by the disk itself, and in the star-disk interaction region.
We cannot claim to have a very complete understanding of the dynamics
of this region, however, until we understand (1) what is the
width of the region of the disk threaded by stellar flux?,
(2) what is the role of outflows in removing angular momentum and
driving the dynamics of this region?, and (3) what is the global geometry
of the stellar and disk magnetic fields?

It is likely that numerical simulations will prove essential in
sorting out many of these issues.  To date, studies of
axisymmetric problems have been reported by a number of
authors.  Most of the simulations produce
outflows, and some produce polar cap accretion (as predicted
by the ``working hypothesis").  However, nonaxisymmetric
simulations are going to be necessary in order to model
the MHD turbulence, dynamo action, angular momentum transport,
and anomalous diffusivity that control the star-disk interaction.

\begin{acknowledgments}
This work was supported by the NSF
through grant AST-9528299.
I thank Steve Balbus, John Hawley, Kristen Miller, Eve Ostriker,
and John Wang for their contributions to much of the work presented
here.
\end{acknowledgments}

\end{document}